\documentstyle[aps,prb,floats,epsf,twocolumn]{revtex}
\fussy
\begin{document}
\wideabs{
\draft
\title{TUNNELING MEASUREMENTS OF THE COULOMB PSEUDOGAP IN A
TWO-DIMENSIONAL ELECTRON SYSTEM IN A QUANTIZING MAGNETIC FIELD}
\author{E.V.~Deviatov, A.A.~Shashkin, V.T.~Dolgopolov}
\address{Institute of Solid State Physics, Chernogolovka, Moscow
District 142432, Russia}
\author{W.~Hansen}
\address{Institut f\"ur Angewandte Physik der Universit\"at Hamburg,
Jungiusstr. 11, 20355 Hamburg, Germany}
\author{M.~Holland}
\address{Department of Electronics and Electrical Engineering,
University of Glasgow, Glasgow G12 8QQ, United Kingdom}
\maketitle

\begin{abstract}
We study the Coulomb pseudogap for tunneling into the two-dimensional
electron system of high-mobility (Al,Ga)As/GaAs heterojunctions
subjected to a quantizing magnetic field at filling factor $\nu\leq
1$. Tunnel current-voltage characteristics show that for the double
maximum observed in the tunnel resistance at $\nu\approx 1$ the
pseudogap is linear in energy with a slope that depends on filling
factor, magnetic field, and temperature. We give a qualitative
account of the filling factor dependence of the pseudogap slope and
we confirm the recently reported appearance of another relaxation
time for tunneling at $\nu\approx 1$. For the tunnel resistance peaks
at $\nu=1/3$ and 2/3 a completely different behaviour of the
current-voltage curves is found and interpreted as manifestation of
the fractional gap.
\end{abstract}
}

\section{Introduction}

The integer quantized Hall effect in a two-dimensional (2D) system is
understood to be a single particle effect. On the other hand it is
now well known that the charge injection by tunneling into the 2D
system at quantizing magnetic fields is very sensitive to many
particle effects. Two kinds of tunneling experiments are possible:
tunneling into the edge (so-called lateral tunneling, e.g.,
\cite{chang,gray,irmer}) and into the bulk (or vertical tunneling
\cite{ashoo,ashoo1,eis1,chan,brown,dolgop1,chan1}) of the 2D system.
It is the vertical tunneling experiments that are a powerful tool for
investigating the electron spectrum of 2D systems. In a pioneering
paper \cite{ashoo} it was shown that the tunneling between a 2D
electron gas and a 3D metallic layer reveals both gaps in the energy
spectrum at integer or fractional filling factors originating from
magnetic field quantization and a Coulomb pseudogap that is pinned to
the Fermi level of the 2D system. The former correspond to cusps in
the dependence of the 2D system energy on electron density and can be
directly measured using alternative methods with or without vertical
tunneling \cite{eis,dolgop} while the latter is seen directly only in
vertical tunneling measurements
\cite{ashoo1,eis1,chan,brown,dolgop1,chan1}. In the work of
Ref.~\cite{ashoo1} the pseudogap was found to persist in a wide range
of magnetic fields, depending weakly on filling factor. Alternative
studies of tunneling between two high-mobility 2D electron systems
pointed to the existence of a pseudogap with exponentially small
tunnel density of states in the extreme magnetic quantum limit
\cite{eis1}. As long as the experiments of Refs.~\cite{ashoo,ashoo1}
and Ref.~\cite{eis1} were performed on samples of very different
quality and in the low or high magnetic field limit, respectively, it
was unclear whether the gap observed by both experimental groups had
the same physical origin. Recently it has been suggested that in all
the above mentioned experiments the same pseudogap was investigated
because all of the obtained results can be reproduced on one sample
\cite{chan}; particularly, at $\nu<1$ the tunnel density of states in
the pseudogap of Ref.~\cite{chan} increases linearly at small and far
more rapidly at larger distances from the Fermi energy. Quite an
unexpected and peculiar result was reported in Ref.~\cite{brown} that
at filling factor $\nu=1/2$ the width of the pseudogap for tunneling
between two equivalent 2D electron layers such as in Ref.~\cite{eis1}
is roughly proportional to magnetic field.

A drastic increase of the pseudogap in the vicinity of filling factor
$\nu=1$ manifested by a camel-back structure of the tunnel resistance
was observed for tunneling from a 3D electron system into a
high-mobility 2D electron gas \cite{dolgop1}. This observation has
been confirmed recently by time-resolved tunneling experiments of
Ref.~\cite{chan1} where a corresponding camel-back structure was
found in the relaxation time of the tunnel current at filling factors
close to $\nu=1$. In that work, however, attention has been paid to
the observation of two strikingly different relaxation times, whereas
the camel-back shape is not discussed. Measurements of
current-voltage characteristics at $\nu\approx 1$ showed that the
pseudogap is roughly linear in energy near the Fermi level
\cite{dolgop1}.

Theoretical models have been developed for the tunneling density of
states in a 2D system in both the metallic and the insulating phase.
In the absence of magnetic field, in a 2D metal with weak random
potential the electron-electron interaction is believed to lead to a
logarithmic correction that reduces the single-particle density of
states at the Fermi energy \cite{alt} whereas in a 2D insulator with
strongly localized electrons the tunnel density of states vanishes at
the Fermi level $\varepsilon_F$ linearly with energy
$D=(2\kappa^2/\pi e^4)|\varepsilon -\varepsilon_F|$, where $\kappa$
is the dielectric permeability and $e$ is the electron charge
\cite{shk}. The problem of vertical tunneling at zero temperature in
quantizing magnetic fields has been tackled in many recent
publications. For the metallic state these predict a Coulomb
pseudogap with exponentially small density of states near
$\varepsilon_F$ \cite{ef,hat,pl,lev}. A similar result was obtained
for the Wigner crystal insulating state \cite{kin} and the insulating
state of weakly disordered 2D systems at high filling factors
\cite{al}. According to Refs.~\cite{ef,hat,kin}, the pseudogap should
scale with the average interelectron distance. Finally, for the
insulating phase with strong disorder the Coulomb-interaction-induced
pseudogap is expected to be linear in energy with a background
density of states at the Fermi energy \cite{mac}

\begin{equation}
D(\varepsilon)=D_F+\alpha |\varepsilon-\varepsilon_F|,\label{eq1}
\end{equation}
where the factor $\alpha$ is a bit different from the constant
$\alpha_0=2\kappa^2/\pi e^4$ predicted in Ref.~\cite{shk}. The linear
dependence (\ref{eq1}) survives even in the presence of a screening
metallic electrode at distances from the 2D system comparable to the
average interelectron distance \cite{pik}.

The above theories give a qualitative account of all experimental
results on the Coulomb pseudogap except as follows: (i) the absence
of scaling of the pseudogap with average interelectron distance
\cite{brown}; (ii) the drop of the coefficient $\alpha$ with magnetic
field \cite{chan}; and (iii) the pseudogap behaviour near filling
factor $\nu=1$ \cite{dolgop1,chan1}.

Here, we study in detail the 3D -- 2D tunneling for high-quality 2D
electron systems in quantizing magnetic fields at $\nu\leq 1$. Except
for the tunnel resistance peaks at $\nu=1/3$ and 2/3, current-voltage
($I-V$) characteristics correspond to a pseudogap that is linear in
energy. For the vicinity of $\nu=1$, where the tunnel resistance
exhibits a double maximum, we analyze the behaviour of the pseudogap
with changing filling factor, magnetic field, and temperature.
Furthermore, we present evidence of the existence of another
relaxation time for tunneling. At filling factor $\nu=1/3$ and 2/3
very distinctive $I-V$ characteristics are observed that are likely
to reflect the fractional gap.
\begin{figure}[t]
\epsfxsize=\columnwidth
\epsffile{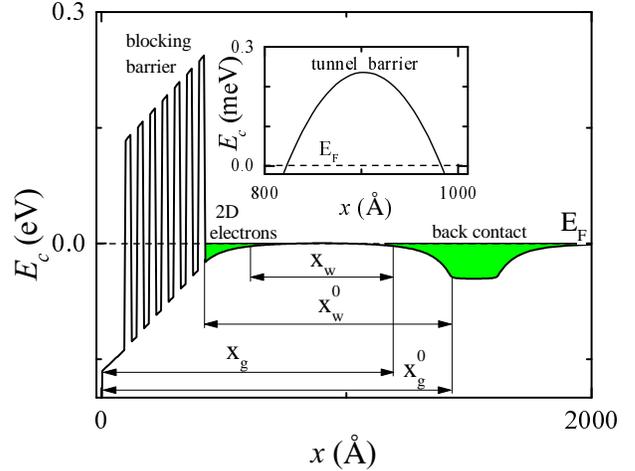}
\caption{Calculated band diagram of the sample at $V_g=0.8$~V. The
$x$ coordinate is counted from the gate. Also shown is a close-up
view of the tunnel barrier region.\label{sample}}
\end{figure}

\section{Samples and experimental technique}

Our samples are metal-insulator-semiconductor (Al, Ga)As/GaAs
heterojunctions with high mobility that contain, apart from a
metallic gate on the front surface, a highly doped ($4\times 10^{18}$
cm$^{-3}$ Si) layer with thickness 20~nm in the bulk of GaAs. This
layer remains well-conducting even at very low temperatures and
serves as a back electrode. The four samples are prepared from two
wafers grown on different MBE machines. The epitaxially grown layer
sequence of the samples and the calculated behaviour of the
conduction band bottom are shown in Fig.~\ref{sample}. A blocking
barrier between the gate and the 2D electrons is formed by a
short-period GaAs/AlAs superlattice capped by a thin GaAs layer. A
wide but shallow tunnel barrier between the back electrode and the 2D
electron system is created by the weak residual p-doping of the GaAs
layer. The density of the 2D electrons is controlled by the gate
voltage $V_g$ applied between the back electrode and the front gate,
because electron transfer across the tunnel barrier brings the
electron gas in the back electrode and the 2D electron system into
equilibrium. The front gate and the heterojunction barrier are
separated from the back electrode by the distances $x_g^0=142$~nm,
$x_w^0=100$~nm in wafer A and $x_g^0=142.4$~nm, $x_w^0=100$~nm in
wafer B. The gate area is equal to $A=8700$~$\mu$m$^2$ for sample A1,
800~$\mu$m$^2$ for samples A2 and A3, and 3300~$\mu$m$^2$ for sample
B.

The gate voltage is modulated with a small {\it ac} voltage so that
an {\it ac} current is excited through the device. From the real and
imaginary components of the current we derive information on both the
thermodynamic density of states and tunnel resistance between the 2D
electron system and back electrode. For the case of linear
current-voltage dependences one obtains \cite{ashoo1,dolgop1}

\begin{equation}
\frac{I}{V}=\omega\left(\frac{\omega\tau+i}{1+\omega^2\tau^2}
(C_{low}-C_{high})+iC_{high}\right),\label{eq2}\end{equation}
where $\omega/2\pi$ is the {\it ac} voltage frequency, $C_{low}$ and
$C_{high}$ are the low and high frequency limits of the device
capacitance, and the relaxation time $\tau$ is equal to

\begin{eqnarray}
\tau=&R&_{tun}(C_{low}-C_{high})\left(\frac{x_g}{x_w}\right)^2,
\nonumber\\
&R&_{tun}=\frac{\tau_{tun}}{AD_Se^2}=\frac{\rho_{tun}}{A},\label{eq3}
\end{eqnarray}
where $R_{tun}$ ($\rho_{tun}$) is the tunnel resistance
(resistivity), $\tau_{tun}^{-1}$ is the attempt frequency, $D_S$ is
the single-particle density of states, and the distances $x_g,x_w$
replace $x_g^0,x_w^0$, taking into account actual electron density
distributions in the $x$ direction (Fig.~\ref{sample}). In the low
frequency limit, the capacitance $C_{low}$ reflects the thermodynamic
density of states \cite{dolgop2}, and the real current component is
proportional to $R_{tun}$. In this limit, nonlinear tunnel $I-V$
characteristics are extracted from the measured $\mbox{Re }I$ and
$\mbox{Im }I$ using the relations for the voltage and current across
the tunnel barrier

\begin{equation}
V_{tun}=\frac{C_{low}\mbox{Re }I}{\omega (C_{low}-C_{high})^2}\left(
\frac{x_w}{x_g}\right)^2,\mbox{ }I_{tun}=\mbox{Im }I.\label{eq4}
\end{equation}
We note that $eV_{tun}$ is defined as the difference of the
electrochemical potentials across the tunnel barrier.

The measurements are performed using a standard lock-in technique in
the frequency interval between 3~Hz and 2~kHz at temperatures between
30 and 880~mK and magnetic fields up to 16~T. The amplitude of the
{\it ac} voltage across the sample is in the range 0.2 -- 8~mV. In
the analysis of nonlinear $I-V$ characteristics we take into account
that it is the first Fourier harmonic of the {\it ac} current which
is measured experimentally.

\begin{figure}
\epsfxsize=\columnwidth
\epsffile{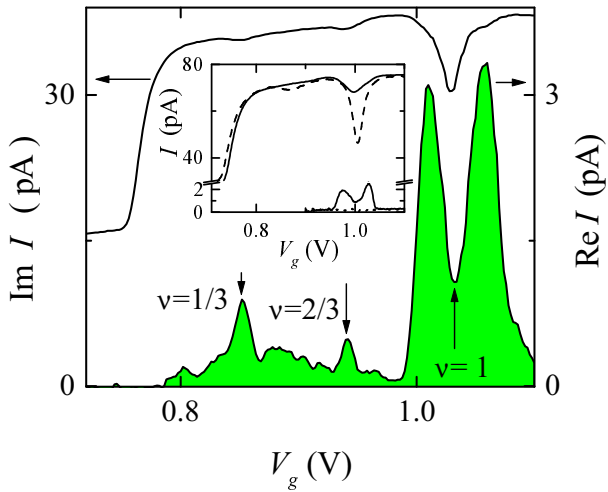}
\caption{Real and imaginary current components as a function of gate
voltage at $T=30$~mK in a magnetic field of 14~T for sample A1;
$V=4.2$~mV, $\omega/2\pi=73$~Hz. The inset compares the experimental
traces at two magnetic fields of 6.9 (dashed lines) and 13.7~T (solid
lines) for sample B; $T=60$~mK, $V=2.1$~mV,
$\omega/2\pi=920$~Hz.\label{trace}}
\end{figure}

A typical experimental trace in the low frequency limit is presented
in Fig.~\ref{trace}. The imaginary current component reflects the
thermodynamic density of states with minima at integer and fractional
fillings and is used to extract the gate voltage dependence of the
electron density \cite{dolgop,dolgop2}. This point is of importance
because lateral transport experiments are principally impossible in a
2D system that is shunted by a 3D back electrode through a tunnel
barrier. The real current component displays a background signal
weakly dependent on filling factor, which was studied previously in
Refs.~\cite{ashoo,ashoo1,eis1,chan,brown}, alongside with a
camel-back structure at $\nu\approx 1$ \cite{dolgop1,chan1} as well
as peaks at $\nu=1/3$ and 2/3. In principle, such additional
structures may be caused by a possible admixture of lateral
transport: the small dissipative conductivity at integer fillings
does not allow tunneling measurements between two identical 2D
electron sheets \cite{eis1,brown}. If the 2D electron system is
charged from the 3D back electrode, there are no restrictions to the
filling factors at which the tunneling experiments can be performed
provided the 2D system and the tunnel barrier are homogeneous. In the
case of an inhomogeneous tunnel barrier the in-plane transport may
contribute significantly to measured values. We argue that the maxima
observed here in the real component of the current close to $\nu=1$
do not reflect lateral transport effects in the 2D system: firstly,
at the same temperature for filling factor $\nu=2$ no peaks are
observed in the real current component although the dissipative
conductivity is expected to be smaller than the one at $\nu=1$ (inset
to Fig.~\ref{trace}); secondly, the behaviour discussed below, in
particular, the frequency, temperature, and magnetic field
dependences of the active current component as well as the behaviour
of $I-V$ curves are inconsistent with the assumption of in-plane
transport; thirdly, very similar data at $\nu=1$ have been obtained
on samples with a different design of the tunnel barrier
\cite{chan1}.

Because on one hand, the dissipative conductivity at fractional
minima is expected to be higher than at $\nu=1$ and, on the other
hand, the amplitudes of the observed peaks at $\nu=1/3$ and 2/3 are
comparable to the value of $\mbox{Re }I$ at $\nu=1$
(Fig.~\ref{trace}), we conclude that the peaks observed at fractional
filling factors are not due to lateral transport either.

\section{Experimental results}

\begin{figure}
\epsfxsize=\columnwidth
\epsffile{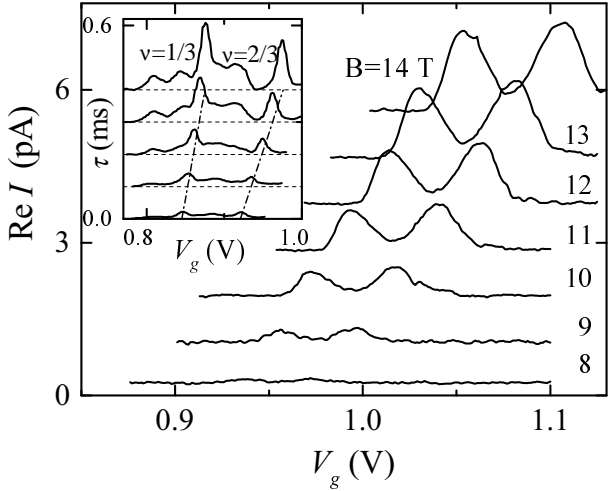}
\caption{A set of the gate voltage dependences of the real current
component near $\nu=1$ at different magnetic fields on sample A1;
$T=30$~mK, $V=0.89$~mV, $\omega/2\pi=93$~Hz. The lines are shifted
vertically for clarity. Inset: a similar set for $\nu<1$ at
$B=12,13,14,15,16$~T with the ordinate axis converted into the
relaxation time.\label{A}}
\end{figure}

\begin{figure}
\epsfxsize=\columnwidth
\epsffile{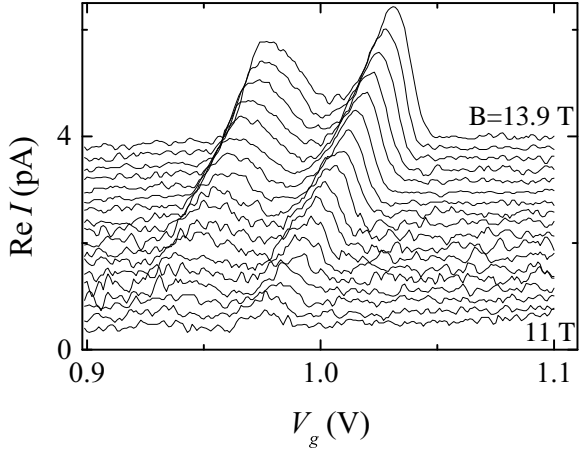}
\caption{The same as in Fig.~\protect\ref{A} for sample B with
magnetic field increment equal to 0.17~T; $T=60$~mK, $V=2.1$~mV,
$\omega/2\pi=920$~Hz.\label{B}}
\end{figure}

The genesis of the camel-back structure centered at $\nu=1$ with
magnetic field is shown in Figs.~\ref{A} and \ref{B} for the samples
of both wafers. This structure emerges at a lower magnetic field for
the higher quality 2D electron system of wafer A as indicated by the
presence (absence) of tunnel resistance peaks at fractions $\nu=1/3$
and 2/3 for wafer A (B). As seen from the figures, the structure has
a double-peak shape over the entire range of magnetic fields.

\begin{figure}[t]
\epsfxsize=\columnwidth
\epsffile{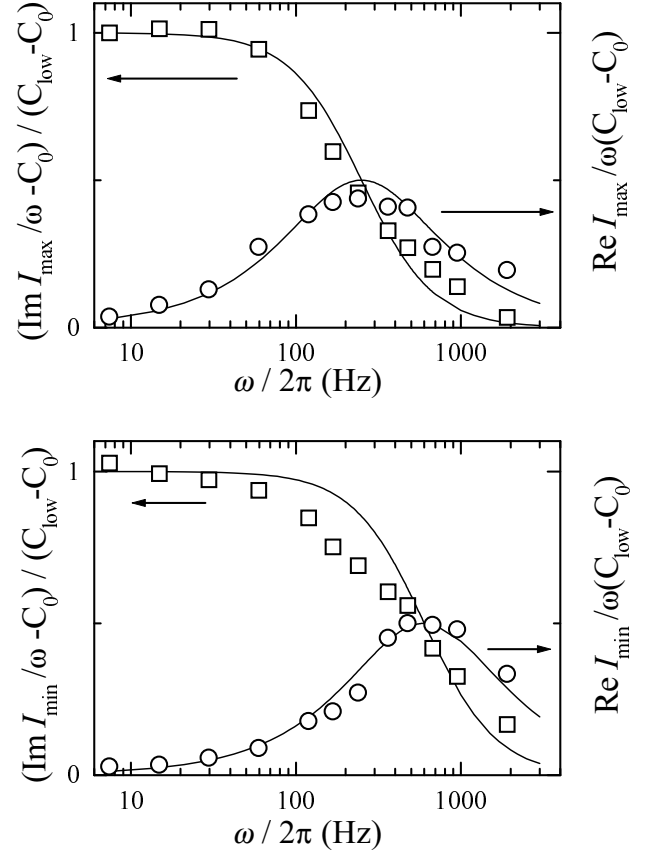}
\caption{Dependence of both current components on frequency at a
maximum (top) and minimum (bottom) of $R_{tun}$ at $\nu\approx 1$ for
sample A1; $B=10$~T, $T=30$~mK. The ordinate axis is normalized to
compare the data with the fit (solid lines) by
Eq.~(\protect\ref{eq2}) with the two parameters $\tau$ and $C_0$ as
described in the text.\label{w}}
\end{figure}
The frequency dependences of both current components measured at a
maximum and at the minimum of the tunnel resistance at $\nu\approx 1$
are presented in Fig.~\ref{w}. Since at fixed modulation voltage $V$
the tunnel current $I_{tun}$ would rise with frequency, we reduce $V$
to keep $I_{tun}$ fixed and thus avoid possible influence of
nonlinearities in this measurement. The formula (\ref{eq2}) fits well
the data points if the capacitance $C_{high}$ in Eq.~(\ref{eq2}) is
replaced by a fitting parameter $C_0> C_{high}$. This implies the
presence of at least two tunneling channels with strongly different
relaxation times: the parameter $\beta=(C_{low}- C_0)/(C_{low}-
C_{high})$ is a weight of the tunneling channel with the highest
tunnel resistivity $\rho_{tun}$ (and $\tau$) so that $\beta A$
describes an "effective area" in the sample, which can be expected to
be cluster-like as inferred from the absence of lateral transport.
The corresponding maximum of the real current component as a function
of frequency is proportional to $\beta$. In the low frequency limit
discussed below, the tunnel current through the so-introduced
effective area is also proportional to $\beta$, i.e., one should
replace $I_{tun}$ in Eq.~(\ref{eq4}) by $I_{tun}=\beta\mbox{Im }I$,
whereas the expressions (\ref{eq3},\ref{eq4}) for $V_{tun}$ and
$R_{tun}$ remain valid since both of these values are related to the
effective area. The fit in Fig.~\ref{w} yields $\beta\approx 1$ and
$\beta\approx 0.6$ for the maximum and minimum of the tunnel
resistance, respectively. We find that these values do not
significantly change with magnetic field. Furthermore, on all samples
the value $\beta$ exhibits a minimum at $\nu=1$ that is similar to
the $\nu=1$ minimum in $C_{low}$ (Fig.~\ref{trace}). We emphasize
that the characteristic double-peak shape persists in $R_{tun}$ (or
$\rho_{tun}$): after dividing the double peak in $\mbox{Re }I$ by
$\beta(C_{low}-C_{high})^2$ for extracting the tunnel resistance
there remains still a minimum at $\nu=1$ with $\approx 30$\% depth.

\begin{figure}[th]
\epsfxsize=\columnwidth
\epsffile{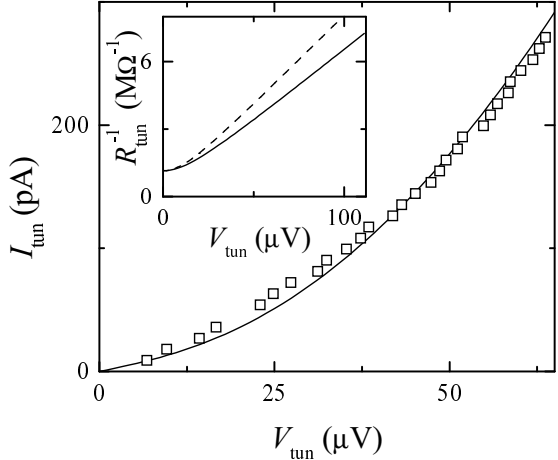}
\caption{Fit using Eq.~(\protect\ref{eq5}) (solid line) of the
experimental $I-V$ characteristic at a maximum of $R_{tun}$ around
$\nu=1$ for sample B at $B=13$~T and $T=60$~mK. The inset displays
this fit in the ($V_{tun},1/R_{tun}$) plane before (dashed line) and
after (solid line) taking the Fourier transform.\label{BIV}}
\end{figure}

\begin{figure}[t]
\epsfxsize=\columnwidth
\epsffile{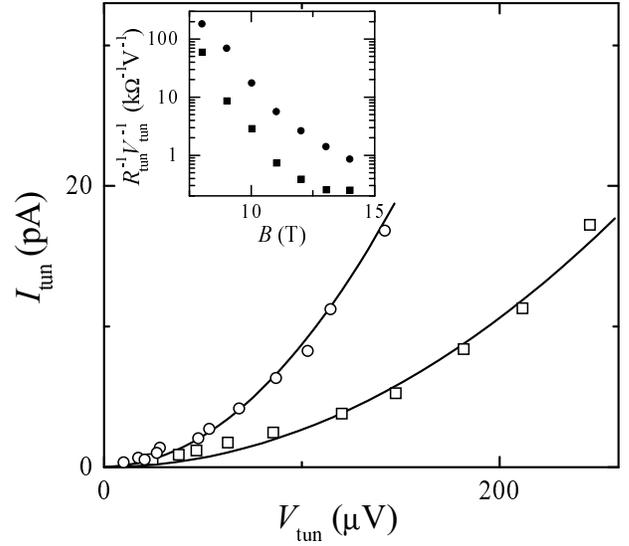}
\caption{Experimental $I-V$ characteristics and their fit (solid
lines) using Eq.~(\protect\ref{eq5}) at a maximum and minimum of
$R_{tun}$ around $\nu=1$ for sample A1 at $B=14$~T at $T=30$~mK.
Inset: behaviour of the pseudogap slope with magnetic field at a
maximum and minimum of $R_{tun}$ near $\nu=1$ for sample A1 at
$T=30$~mK.\label{AIV}}
\end{figure}

The experimental $I-V$ characteristics for the camel-back structure
are depicted in Figs.~\ref{BIV} and \ref{AIV}. At all magnetic
fields, these are parabolic at $eV_{tun}>k_BT$ and linear at
$eV_{tun}<k_BT$ as caused by temperature smearing. The parabolic
behaviour of the $I-V$ curves corresponds to a linear pseudogap. To
describe the data we calculate the first Fourier harmonic of the
voltage on the nonlinear element $V_{tun}(I_{tun})$ defined by the
expression

\begin{equation}
I_{tun}=\gamma\int^\infty_{-\infty}D_mD(\varepsilon)[f(\varepsilon-
eV_{tun},T)-f(\varepsilon,T)]{\rm d}\varepsilon,\label{eq5}
\end{equation}
where $\gamma$ is a factor and $D(\varepsilon)$ is given by
Eq.~(\ref{eq1}) with $D_F=0$. The density of states $D_m$ in the back
electrode is assumed to be featureless even in high magnetic fields
because of the low mobility in the highly doped layer, and
$f(\varepsilon,T)$ is the Fermi distribution function. So-calculated
$I-V$ characteristics fit the experiment very well with only one
fitting parameter $\alpha\gamma D_m$ (Figs.~\ref{BIV},\ref{AIV}). For
visualization purpose the solid line from Fig.~\ref{BIV} is drawn in
the coordinates ($V_{tun},1/R_{tun}$) in the inset to Fig.~\ref{BIV}.
Also shown by a dashed line is the corresponding dependence before
filtering out the first Fourier component. One can see the
finite-temperature-induced saturation of $1/R_{tun}$ at
$V_{tun}\rightarrow 0$.

\begin{figure}
\epsfxsize=\columnwidth
\epsffile{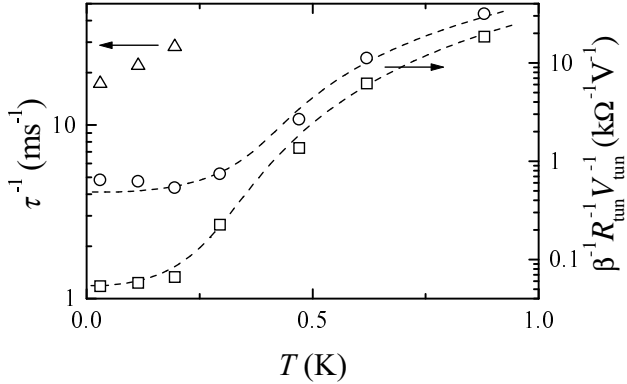}
\caption{Temperature behaviour of the pseudogap slope at a maximum
and minimum of $R_{tun}$ on sample A1 at $\nu\approx 1$ and $B=14$~T.
The dashed lines are guides to the eye. Also shown is the temperature
dependence of the relaxation time for the corresponding tunnel
resistance peak at $\nu=2/3$.\label{BT}}
\end{figure}

Assuming that for a given sample the values $\gamma$ and $D_m$ are
constant, the pseudogap parameter $\alpha$ depends on magnetic field
and temperature in the same way as the experimentally determined
slope of the dependence $R_{tun}^{-1}(V_{tun})$ on the parabolic part
of $I-V$ curves, see the inset to Fig.~\ref{AIV} and Fig.~\ref{BT}.
To our surprise, we find that the value $\alpha$ changes strongly
with both magnetic field and temperature. At high magnetic fields and
low temperatures there is a tendency to saturation of $\alpha$ (that
the electron temperature is not saturated in the low temperature
limit is indicated by the pronounced temperature dependence of the
tunnel resistance at $\nu=2/3$, see Fig.~\ref{BT}) while at low
magnetic fields and high temperatures the pseudogap disappears.

\begin{figure}[t]
\epsfxsize=\columnwidth
\epsffile{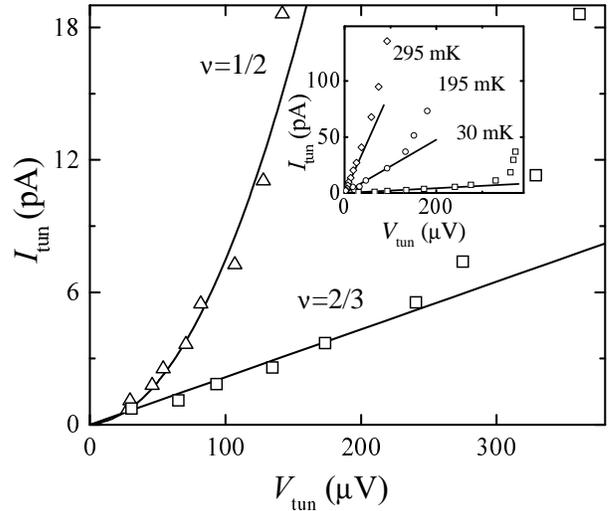}
\caption{Comparison of the experimental $I-V$ curves at $\nu=1/2$ and
$\nu=2/3$ on sample A1 at $B=16$~T and $T=30$~mK. The $\nu=1/2$ data
points are fitted using Eq.~(\protect\ref{eq5}) as described in the
text. The initial interval of the $\nu=2/3$ $I-V$ curve is fitted by
a straight line. The $\nu=2/3$ $I-V$ dependence at different
temperatures is presented in the inset.\label{fr}}
\end{figure}

The active current component at $\nu<1$ converted into $\tau$ by
means of Eq.~(\ref{eq2}) is displayed for different magnetic fields
in the inset to Fig.~\ref{A}. The relaxation time $\tau$ turns out to
be of the same order of magnitude as the one of Ref.~\cite{chan} and
the slow relaxation time of Ref.~\cite{chan1}. Also, the magnetic
field dependence of the background signal is close to that reported
in Refs.~\cite{chan,brown}. In contrast, peaks at filling factors
$\nu=1/3$ and 2/3 were never observed in previous publications, we
believe, because of a problem with lateral transport in
Ref.~\cite{eis1} and too low magnetic fields used in
Ref.~\cite{chan1}. As seen from the figure, the $\nu=2/3$ peak is
less influenced by the background and so it is more suitable for
investigations. The low temperature $I-V$ characteristics for the
background at $\nu=1/2$ and for the $\nu=2/3$ peak are compared in
Fig.~\ref{fr}. While the former is close to parabolic at
$eV_{tun}>k_BT$, the $\nu=2/3$ current-voltage characteristic stays
linear up to much higher voltages until it increases abruptly. As
seen from the inset to Fig.~\ref{fr}, the observed linear region
shrinks with temperature.

\section{Discussion}

In view of the existence of a pseudogap at $\varepsilon_F$ in a 2D
electron system over a wide region of filling factors, our data are
consistent with results of preceding experimental and theoretical
publications. We confirm that the pseudogap is linear in energy for
the background at $\nu<1$ and establish that the linear law is valid
also for the camel-back structure around $\nu=1$. A linear pseudogap
may be anticipated at $\nu\approx 1$ where the localization length is
expected to be small so that the limiting case of electrons localized
on individual impurities is approached \cite{shk,mac}. While the
theory predicts the universal pseudogap slope $\alpha_0=2\kappa^2/\pi
e^4$, we find that the parameter $\alpha$ depends on temperature and
magnetic field so that it saturates at high magnetic fields and low
temperatures. In view of the fact that in this limit the electrons
are best localized we may anticipate that in this limit the classical
value $\alpha_0$ is approached.

The characteristic dependence of the pseudogap parameter $\alpha$ on
filling factor near $\nu=1$ is reflected by the double-peak tunnel
resistivity $\rho_{tun}$ as described above. Particularly, from the
analysis of $I-V$ curves it follows that the slope $\alpha$ reaches a
maximum at $\nu=1$ (Fig.~\ref{AIV}). Comparison of the positions of
the $\mbox{Re }I$ (or $\rho_{tun}$) peaks around $\nu=1$ with the
metal-insulator phase diagram obtained on samples of similar quality
\cite{phd} shows that the peak position is close to the
metal-insulator transition point. Hence, as the filling factor
deviates from $\nu=1$, the pseudogap parameter $\alpha$ decreases and
then passes through a minimum near the metal-insulator transition.

The theoretical models developed heretofore allow one to account
qualitatively for the above behavior of $\alpha$ with filling factor.
According to Refs.~\cite{pl,lev}, for the metallic phase the
tunneling-caused excessive charge should accommodate through the
dissipative conductivity in the 2D plane. The higher the
conductivity, the lower the resulting tunnel barrier because of
faster charge accommodation and, therefore, the pseudogap narrows as
one advances deeper into the metallic phase. As mentioned above, an
exponential pseudogap is expected in the metallic phase at zero
temperature

\begin{equation}
D(\varepsilon)=D_{th}\exp\left(-\ln^2(e^4/\kappa^2K|\varepsilon-
\varepsilon_F|)\right),\label{eq6}\end{equation}
where $D_{th}$ is the thermodynamic density of states and $K$ is the
diffusion coefficient \cite{lev,pol}. This conflict between the
linear and exponential pseudogap can be sorted out as follows. An
idea has been expressed in Ref.~\cite{pol} that, given the average
size $\xi$ of the conducting clusters in the insulating phase, the
dependence (\ref{eq6}) should be replaced in the energy interval
$|\varepsilon- \varepsilon_F| <U_c=e^2/\kappa\xi$ by $\alpha
|\varepsilon- \varepsilon_F|$ with $\alpha=D(\varepsilon_F
+U_c)/U_c$. In agreement with our finding, the so-defined $\alpha$
enhances at $\nu\rightarrow 1$ because the correlation length $\xi$
decreases when going deeper into the insulating phase. Following the
approach of Ref.~\cite{pol}, for actual samples where the correlation
length is always restricted one expects a linear pseudogap in the
close vicinity of $\varepsilon_F$ in the metallic phase as well. This
allows reconciliation of the data at $\nu<1$ on the linear pseudogap
(Ref.~\cite{chan} and the present paper) and the exponential one
\cite{eis1,brown} as obtained in very different ranges of tunnel
voltages.

Still, the temperature and magnetic field dependences of $\alpha$ as
well as the prominence of the filling factor $\nu=1$ with respect,
e.g., to $\nu=2$ cannot be explained by existing theories. In the
vicinity of $\nu=1$ our data indicate a decrease of the effective
area that is accompanied by the emergence of another, shorter
relaxation time related to the "remaining area", which is in
agreement with results of Ref.~\cite{chan1}. As mentioned above, we
preclude the possibility that the factor $\beta$ describes a
macroscopic area in the sample with high tunnel resistivity because
this would imply the presence of lateral transport. The origin of the
effect may be different rates for tunneling into the edge and the
bulk of electron islands whose size $\xi$ exceeds by far the magnetic
length. Indeed, with decreasing $\xi$ the conducting clusters break
up into smaller ones so that the fraction of electrons near the
internal "edges" with lower tunnel resistivity should increase
reaching a maximum at $\nu=1$. Another way of explanation is to
invoke spin effects as has been suggested in Ref.~\cite{chan1}.

Intriguingly we find a completely different behavior of the $I-V$
curves at fractional filling factor $\nu=1/3$ and 2/3 where the
tunnel current rises linearly with the voltage up to a critical
voltage that drops as the temperature is increased, see the inset to
Fig.~\ref{fr} and Fig.~41 from Ref.~\cite{qq}. This is likely to
point to a real gap in the 2D spectrum at fractional $\nu$ that
collapses with temperature. We find that at the lowest temperatures
the estimated gap value is close to the value obtained in
Ref.~\cite{qq}. So, the gaps for tunneling at $\nu=1/3$ and 2/3 are
not manifested by double-peak structures such as at $\nu=1$ and have
a different energy dependence whilst the gaps in the thermodynamic
density of states at all of these filling factors look similar
\cite{dolgop,dolgop2}.

\section{Conclusion}

In summary, we have performed vertical tunneling measurements to
investigate the Coulomb pseudogap in a 2D electron system of
high-mobility (Al,Ga)As/GaAs heterojunctions subjected to quantizing
magnetic fields. From the analysis of $I-V$ characteristics it has
been found that the pseudogap is linear in energy for both the
background tunnel resistance at $\nu<1$ and the camel-back structure
at $\nu\approx 1$. The filling factor dependence of the pseudogap
slope $\alpha$ near $\nu=1$ can be explained qualitatively by theory
whereas the observed change of $\alpha$ with temperature and magnetic
field is not yet clear. We give independent confirmation of the
appearance at $\nu\approx 1$ of another, shorter relaxation time for
tunneling. The very different $I-V$ curves found at the $\nu=1/3$ and
2/3 peaks of the tunnel resistance are interpreted as manifestation
of the fractional gap. Its estimated value and temperature behaviour
agree well with results of previous studies.

\acknowledgements

We would like to thank P.~Kopietz, K.V.~Samokhin, and A.V.~Shitov for
valuable discussions and G.E.~Tsydynzhapov and I.M.~Mukhametzhanov
for technical assistance. Furthermore, we are very grateful to
D.~Schmerek and H.J.~Klammer for sample preparation. The band diagram
was calculated using a programme by G.~Snider. This work was
supported in part by the Russian Foundation for Basic Research under
Grant No.~97-02-16829, the Programmes "Nanostructures" under Grant
No.~97-1024 and "Statistical Physics" from the Russian Ministry of
Sciences, and the Deutsche Forschungsgemeinschaft via the
Graduiertenkolleg "Physik nanostrukturierter Festk\"orper".

\end{document}